# END-TO-END LSTM BASED ESTIMATION OF VOLCANO EVENT EPICENTER LOCALIZATION

Néstor Becerra Yoma[1], Jorge Wuth[1], Andrés Pinto[1], Nicolás de Celis[1], Jorge Celis[1], Fernando Huenupan[2]

[1]Speech Processing and Transmission Laboratory, Electrical Engineering Department, University of Chile, Santiago, Chile

[2]Department of Electrical Engineering, Universidad de La Frontera, Av. Francisco Salazar 01145. Temuco, Chile

## ABSTRACT

In this paper, an end-to-end based LSTM scheme is proposed to address the problem of volcano event localization without any a priori model relating phase picking with localization estimation. It is worth emphasizing that automatic phase picking in volcano signals is highly inaccurate because of the short distances between the event epicenters and the seismograph stations. LSTM was chosen due to its capability to capture the dynamics of time varying signals, and to remove or add information within the memory cell state and model long-term dependencies. A brief insight into LSTM is also discussed here. The results presented in this paper show that the LSTM based architecture provided a success rate, i.e., an error smaller than 1.0Km, equal to 48.5%, which in turn is dramatically superior to the one delivered by automatic phase picking. Moreover, the proposed end-to-end LSTM based method gave a success rate 18% higher than CNN.

## 1. INTRODUCTION

### 1.1 Volcano event location with P and S wave picking.

Locating sources of volcano-seismic signals is important for monitoring active volcano and understanding volcanic processes (Morioka, Kumagai & Maeda, 2017). Classical estimation of source seismic events is based on phase picking (Pardo, Garfias & Malpica, 2019). The most accurate method of phase selection is the visual inspection of the records by experts, who take advantage of the global information of each event, local clues for phase detection and comparison with records from other stations. However, this activity is highly time demanding.

The location of seismic events and, in particular, volcanic tectonic (VT) events, can be determined by triangulation if the distances of the event to at least 3 stations are known. The distances to each station can be determined, in principle, using the differences in the arrival time of the P and S waves combined with a propagation model (Stein & Wysession, 2009). In the case of VT events, it is possible to distinguish the arrival of P and S waves. As far as we know, the problem of the location of VT events with automatic phase picking has not been addressed or employed in the literature of exhaustively.

As mentioned above, given an event recorded at a station, it is possible to estimate the distance from the epicenter of the event to the sensor using the information about the arrival times of the P and S waves at the station. In a simplified case, a linear relationship between the difference in time between the arrival of the P and S waves and the epicenter-station distance can be applied, approximating the propagation speed to a constant. For this case, which is the simplest one, a circumference is drawn at each station where the event was detected, where the radius is proportional to the P-S phase difference, and the intersection of these circles would be the location of the event. The more sensors that record the event, the better the location accuracy. This model does not consider the speed changes caused by the different materials that make up the propagation medium, which in turn compromises the method accuracy. More complex propagation models have been included in the last decades and the epicenter-station distance is no longer represented by a circle (Lomnitz, 1977; Fatolazadeh et al., 2020; Bondár et al., 2004). Also, due to the low energy of the signals, sensors need to be positioned in the vicinity of the volcano events to detect them.

Unlike VT events, the problem of automatic phase selection in earthquake tectonic seismic events has been widely addressed in the literature using STA / LTA detectors, kurtosis and asymmetry as shown in (Ross et al., 2014; Álvarez et al., 2013; F. Tan et al., 2019). In recent years, much of the research on automatic phase detectors has focused on deep learning due to the excellent results obtained such as those shown in (Zhu et al., 2019a; Mousavi et al., 2020; Tsai et al., 2018).

The accuracy of the proposed methods for automatic phase selection often depends on the database to which they are applied. In general, their performance depends on a high SNR to be able to discriminate the P and S waves. Furthermore, few of them are applicable to VTs (García, 2019). This may be because, in contrast with tectonic seismology, "when the earthquake source occurs in a volcano, the shallowness, proximity of the source, lower magnitude, and complex propagation path, among other factors, make the S-phase detection more difficult." (García, 2019). Moreover, according to (Gasparini, Scarpa & Aki, 2012), "the structure inside a volcano is, probably, the most complex subject seismologists have ever encountered on Earth. It is extremely heterogeneous, anisotropic and absorptive with irregular interfaces and topography including cracks of all sizes and orientation. The source processes are also much more complicated than usual tectonic earthquakes, because they involve additional dynamics of gas and fluid magma in generating seismic signals, to be deciphered by seismologists." As a result, localizing VT events is an open and challenging problem.

## 1.2 Volcano event location based on machine learning

To the best of our knowledge, the problem of VT event localization with machine learning has not been addressed in the literature exhaustively. In contrast, several methods have been proposed in earthquake seismology. In (Perol, Gharbi & Denolle, 2018) a technique based on CNN is presented using the time domain signals corresponding to the three channels of a single station. The location problem is approached as a classification one dividing the study area into six groups of approximately 100 km$^2$ each. They achieve a correct identification of clusters in 74.5% of the cases. In (Kriegerowski, M. et al 2019) a localization method of seismic events with CNN using signals from multiple stations with three channels each was proposed. The employed CNN architecture combines the time domain signals of the stations and obtains a prediction error of less than 200 m in 86% of the cases. In (Mosher & Audet, 2020), machine learning was used to detect and locate seismic events in the northern area of the San Andreas fault by using 76 stations separated by 162km in average. For this, only the vertical component of the seismographic records was employed. A multilayer perceptron neural network was trained, which obtained a 97.4% precision in detection. However, a grid was adopted for earthquake location, which carries a minimum intrinsic error of 4.33 km. In (Zhang, et al., 2020) small earthquakes induced by oil and gas industries are located with low SNR. A 22-layer CNN network was employed, to which the signals are input in the time domain as images. The network was trained with 1013 tagged events, where additional data augmentation was added. The average location error was equal to 4.9 km.

Various machine learning techniques have been applied to analyze volcanic seismic events. Approaches include support vector machines (SVM) (Masotti et al., 2006), hidden Markov models (HMM) (Cassisi et al., 2016; Quang et al., 2015), Gaussian mixture models (Cortés et al., 2014), and multilayer perceptron (MLP) (Esposito et al., 2018; Esposito et al., 2013). Deep learning approaches have also been applied and shown to outperform MLP, SVM, and random forest for the classification of volcanic seismic events (Bicego, Londoño-Bonilla & Orozco-Alzate, 2015). According to (Bicego, Londoño-Bonilla & Orozco-Alzate, 2015), the pre-training initialization of the DNN model was effective in supporting the training process. However, according to the authors, the resulting DNN cannot be applicable from one volcano to another one because of the intrinsic volcano process and event variability. In (Malfante et al., 2018) an analysis of the state-of-the-art in machine learning techniques for volcanic-seismic signals is presented. The problem is divided into stages of detection and classification. Even though the results are promising, the authors point out that the detection and classification of overlapped volcanic-seismic events in continuous records is a difficult task, particularly during an eruption. As can be seen, in volcanology deep learning has been employed mainly for the detection and classification of volcanic events and the epicenter location problem has been rather neglected. Some of the architectures used and their applications are described below.

Convolutional neural networks (CNN) have been used with great success for image recognition (Krizhevsky, Sutskever & Hinton, 2012; Simard, Steinkraus & Platt, 2003) and have also been applied in the field of volcanology. A CNN contains layers that convert the pixel intensities of an image and send them to the next layer. The last layer of the CNN generates the probabilities that the object being evaluated will fall into each class. In (Anantrasirichai et al., 2019), a CNN-based system is presented to detect volcano deformation in satellite images. The authors demonstrate that training with synthetic examples can enhance the ability of CNNs to detect volcano deformation. CNN has also been used in (Lara et al., 2021) for the detection and classification stages of an automatic recognition system for volcanic micro-earthquakes of the Cotopaxi Volcano. The system led to accuracies of 99% and 97% in the detection and classifications procedures, respectively.

## 1.3 Long Short-Term Memory neural networks

Recurrent neural networks (RNN) can translate input sequences into output sequences (LeCun, Bengio & Hinton, 2015; Schmidhuber, 2015) making use of temporal modeling capabilities. This makes the RNN suitable for processing volcanic-seismic events from seismic data, as this is a sequential problem involving complex and highly dimensional dynamic signals. In (Titos, et al, 2019) RNN was employed for automatic detection and classification of volcano seismic signals. Surprisingly, RNNs have not been explored exhaustively to address the problem of volcano event epicenter localization.

For very long sequences, the gradient may disappear or explode when computing the derivatives of the error. This imposes restrictions when capturing long-term dependencies (Schmidhuber, 2015; Pascanu, Mikolov & Bengio, 2013). To overcome this problem, a recurring alternative known as long short-term memory (LSTM) was proposed in (Hochreiter & Schmidhuber, 1997).

LSTM has been used for the recognition system of volcanic-seismic events (Titos et al., 2018; Canario et al., 2020), identification of P waves in tectonic seismic data (Zhu et al, 2019b) and prediction of seismic events (Nicolis, Plaza & Salas, 2021). In this paper, LSTM is employed to estimate the epicenter of volcano events by exploring the LSTM inner structure to capture the relevant characteristics of the wave signal to achieve this goal, i.e. the occurrence of P and S waves, on end-to-end basis.

In this section we will provide an insight into LSTM neural networks to justify its use to address the problem of volcano event epicenter estimation without phase picking. Many papers just use machine learning schemes without providing a justification based on theoretical description of the chosen technique. Also, an introduction on LSTM may be useful to those readers that are not familiar with this recurrent neural network. As discussed above, automatic phase picking in volcano events is particularly difficult because the distance between the epicenter and the stations is small. In the problem considered here, we need the recurrent network to capture the time dynamics of volcano signals, and to detect implicitly the P and S waves and the transition between them. LSTM networks can remove or add information within the memory cell state, regulated with several gates that allow information to flow over time and model long-term dependencies. An LSTM unit includes a memory cell, an input gate, an output gate, and a forget gate. The memory cell is used to store values in time intervals -the cell state- that modulates the hidden state/output. The gates control how these values are updated over time. By doing so, an LSTM can process sequential data and retain states over time. A graphical representation of the LSTM cell is shown in Fig. 1. The input, the hidden state (or output) and the cell state are related through the input gate, forget gate and output gate as follows:

$$i_t = \sigma(W_i Z_t + U_i h_{t-1}) \tag{1}$$

$$f_t = \sigma(W_f Z_t + U_f h_{t-1}) \tag{2}$$

$$o_t = \sigma(W_o Z_t + U_o h_{t-1}) \tag{3}$$

$$\tilde{c}_t = \tanh(W_c Z_t + U_c h_{t-1}) \tag{4}$$

$$C_t = f_t \odot c_{t-1} + i_t \odot \tilde{c}_t \tag{5}$$

$$h_t = o_t \odot \tanh(C_t) \tag{6}$$

where $Z_t$, $h_t$, $C_t$ are the exogen input, the hidden state or output and the cell state vectors at frame t, respectively; $i_t$, $f_t$, and $o_t$ are the input gate, the forget gate and the output gate vectors at frame t, respectively; $W_i$, $W_f$, $W_o$ and $W_c$ are gate weight matrices that corresponding to input $Z_t$; $U_i$, $U_f$, $U_o$ and $U_c$ are the weight matrices corresponding to the previous frame output; and, $\odot$ denotes the pointwise product operator between two vectors. To understand how the LSTM works, the computation of $h_t$, $C_t$ at frame t will be discussed. The network information up to frame t is contained in $h_{t-1}$ and $C_{t-1}$. The information in the cell state can be understood as a vector of weights that modulate the estimation of output $h_t$, which in turn is the endogen input at frame t+1. In this sense, "forgetting" or "adding" information is modeled as weighting with values less than 1 or adding values greater than 0, respectively. To achieve these dynamics, $h_{t-1}$ and $Z_t$ are processed to generate vectors that are operated pointwise with the cell state $C_{t-1}$. Forget gate generates a vector with values less than 1, according to the sigmoidal activation function in (2), which weights $C_{t-1}$ by means of the pointwise product operator as can be seen in (5). Input gate generates a vector that is combined with a candidate cell state obtained in (4) by using the pointwise product operator. The result is added to $f_t \odot c_{t-1}$ to determine a new cell state, i.e. $C_t$. Finally, the output gate estimates the next hidden state as follows. First, $o_t$ is estimated according to (3). Then, $h_t$ is estimated as in (6) to decide what information the hidden state should carry. The updated new cell state and the new hidden state are then carried over to the next time step.

## 2. LSTM BASED END-TO-END VOLCANO EVENT LOCATION

End-to-end based schemes are becoming a promising practical approach that attempts to deliver good results without a priori models (Liu et al., 2019; Shi et al., 2019). It has provided very competitive results in the field of automatic speech recognition by using only input-output information (Audhkhasi et al., 2017; Li et al., 2020; Zeineldeen et al., 2020). In our case, the motivation to use an end-to-end based approach is to avoid the need to detect the P and S waves to estimate the volcano event epicenter from the station signals. As mentioned above, the distance between the seismic stations and volcanic events is short so the automatic discrimination between P and S waves is a difficult problem. As all the information necessary to locate an epicenter is contained in the seismic waves, we propose a LSTM based method to estimate the epicenter position without making an explicit phase picking of the P and S waves.

### 2.1 Proposed architecture

Recurrent neural networks stand out for their ability to learn sequential patterns or variants over time (Medsker et al., 1999). For this reason, in a seismic location problem where is essential to estimate the arrival time of P and S waves, it is expected that recurrent networks have the ability to solve this problem. However, as mentioned above, ordinary RNNs are difficult to train since they usually run into the vanishing gradient problem. According to (Staudemeyer & Morris, 2019), this problem prevents RNNs from learning sequences longer than 10 steps. To overcome this challenge, as discussed in section 1.3, the LSTM architecture was introduced to preserve important information while discarding the irrelevant one in a time data sequence. By doing so, LSTM should be able to register the arrival time of P and S waves if it is trained to estimate volcano event epicenters. In contrast, it should ignore any signal component, i.e. noise, that is not relevant for the task. The proposed network architecture is shown in Fig. 2. It employs a multiple layers of LSTM units. The LSTM output at the last frame is input to a feed-forward fully connected neural networks whose output corresponds to the latitude and longitude of the estimated epicenter. Comparative experiments with CNN are also provided because these neural networks have been employed quite extensively in volcanology and also in seismology.

In this paper, only the vertical component of the seismic signals recorded by the monitoring stations was used. As mentioned in (Bormann et al., 2013) a P wave can be divided into two main components, one vertical and one horizontal, where the first record of the P wave is obtained from the vertical axis. On the other hand, in (Trnkoczy et al., 2009) it is shown that there is greater noise in the horizontal components because the sediments show special tilting effects. Moreover, if the noise levels are comparable at the three components, using only the vertical component records should be enough.

### 2.2 Pre-processing

To keep the experimental conditions under control, all the recorded event signals were segmented to last seven seconds making sure that there was two seconds before the P signal. Then, each signal was subdivided into windows of 0.5 seconds with a 66% overlap. The LSTM network input corresponds to an M-by-N matrix, where M is the number of frames and equal to 39, and N Is the number of features per frame. The features (denoted by Z) extraction process is depicted in Fig. 3 and is achieved by applying the 64-sample FFT, absolute value and logarithm at each frame. The two lowest frequency bins were removed, which in turn leads to a total number of features per frame equal to 31. Finally, mean and variance normalization was applied to each feature trajectory. The resulting features correspond to input $Z_t$ in (1),(2),(3) and (4).

# 3. EXPERIMENTS

## 3.1 Database

The database was provided by the Southern Andes Volcano Observatory, OVDAS ("Observatorio Volcanológico de los Andes Sur") and it consists of volcanic seismic events from Chillan Volcano, Chile. This volcano is part of Nevados de Chillán and is a large composite stratovolcano complex in the southern Chilean Andes. It is one of the highest-risk volcanoes in the country due to high levels of historic activity and rapid development of human occupation in the area. In January 2020, Chillan volcano presented an eruption that emitted a column of smoke more than 3,500 kilometers high. Later during the year there were more than five explosions in April, May and June. As a result, the original two-kilometer exclusion zone, decreed by the Chilean National Geology and Mining Service, increased to five kilometers to the northwest and three kilometers to the southwest.

The whole database is composed of three datasets: the first one has 753 events from 2015 to 2017; the second subset corresponds to 220 events from 2019; and, and the third one provides 569 events from 2020. The signals are sampled at a rate of 100Hz. All the signals contain VT (volcano-tectonic) events only from 13 stations as shown in Fig. 4. The time of arrival of waves P and S were manually marked by experts.

The results with LSTM and CNN were obtained by dividing the database randomly in training (80%), validation (10%) and testing (10%) subsets. Each event was observed by the same three stations leading to three signals per event in the training, validation and testing subsets. The reference epicenter location required by the supervised training procedure was obtained as follows: first, each event was represented with the signals monitored by all the available stations that ranged from three to ten depending on the event; then, the reference epicenter locations were obtained by processing this information using the specialized Hyposat software (Schweitzer, 2001). This procedure was adopted because the accuracy in the estimation of the event epicenter with Hyposat depended on the number of stations employed per each event.

Also, for comparison purposes, results were obtained with an automatic phase picker. To do so, a band-pass frequency filter was applied to the testing subset. The cut-off frequencies of the filter were 0.8Hz to 10Hz. The P and S waves were picked automatically with a Matlab toolkit (Kalkan, 2021). Then, the same procedure followed to estimate the reference epicenter location as described above was adopted here.

## 3.2 Hyperparameter and architecture optimization

The metric employed in this study corresponds to the Mean Absolute Error (MAE) obtained with respect to the reference of the epicenter location mentioned above. This metric was employed for both, the hyperparameter and architecture optimization, and to compare the proposed technique with CNN and automatic phase picking. The percentage of the event epicenter estimations with error lower than 1Km is called "success rate."

The hyperparameters of the LSTM and the feed-forward fully connected neural networks (see Fig. 2) were optimized with the validation subset. Regarding, the LSTM module, the number of layers and the output dimension were varied from one to six and from 32 to 512 (with power of two steps), respectively. In the case of the fully connected neural network, the number of layers was varied from one to three. Observe that the number of nodes in the first layer of the fully connected neural network is equal to the LSTM output dimension. Also, dropout was applied in this neural network to improve generalization. The learning rate was adjusted in powers of 10 steps ranging from $8 \times 10^{-6}$ to $8 \times 10^{-1}$. The estimated optimal hyperparameters correspond to: number of LSTM layers equal to four; LSTM output dimension equal to 256; and, number of layers in the fully connected neural network equal to two. The optimal learning rate and dropout were equal to $8 \times 10^{-4}$ and 10%, respectively.

The CNN based scheme was also tuned with the validation subset. The best results were obtained using four convolutional layers of 256 3x3 filters for each channel, followed by a three-layer fully connected neural network, where the first two layers have 64 nodes each. As in the case of the proposed LSTM based solution, the last layer of the fully-connected module has two output nodes, i.e. the latitude and longitude of the estimated epicenter. An optimal 30% dropout after each layer was applied.

## 4. DISCUSSIONS

Figure 5 shows the results for event localization with LSTM, CNN and two configurations of the automatic phase picker as described above, i.e. using three and up to ten stations. As can be seen, the best result was obtained with the LSTM based proposed scheme. Using only three sensors as described above, the LSTM based architecture provided a success rate equal to 48.5%. In contrast, the automatic phase picking provided success rates equal to 0.3% and 1% with three and up to ten stations respectively. This comparison demonstrates an overwhelming superiority of LSTM over the automatic phase picking method for volcano event epicenter estimation. As can also be observed in Fig. 5, the proposed end-to-end based LSTM method provides success rate 18% higher than CNN, which in turn delivers a success rate equal to 41%. This result strongly corroborates the initial assumptions that motivated the adoption of LSTM to address the volcano event epicenter problem: first, being a recurrent neural network itself, LSTM provides a suitable framework to capture the dynamics of time series; and second, LSTM allows information to flow over time and model long-term dependencies discarding not relevant temporal information. In contrast, CNN is an architecture that was proposed originally for image recognition. Since then, it has also been employed in fields such as speech processing and seismology. Moreover, CNN was not designed to model long-term dependencies and filter out information that is not pertinent regarding a given task. Notice that both the LSTM and CNN were tuned here in the same conditions.

According to Fig. 6, the accuracy in epicenter estimation decreases when the distance between the volcanic event and the centroid of all the references is larger than 2.5Km, where the reference centroid denotes the mean localization of the training events. Observe that the only exception to this trend corresponds to those events located in the range of 3.5Km-4.0Km from the reference centroid. This finding deserves some discussion. For a radius smaller than 2.5Km there is something like 40% of the training data. In other words, 40% of the training data is allocated in an area of 19.6 Km². Between 2.5Km and 5.5Km there are 50% of the training data approximately in an area of 75.5Km2, which corresponds to a much lower density of training events in average than in the radius smaller than 2.5Km. However, in the range of 3.5Km-4.0Km from the reference centroid there is a higher concentration of training volcano events that corresponds to approximately 20% of the whole training subset. Quite consistently, the localization accuracy with events between 3.5Km and 4.0Km from the reference centroid is as high as with events that are closer to this centroid. This result suggests that the accuracy of the proposed system depends on the density of training data. Actually, Figure 7 shows a clear trend where the lower the training data density, the lower the success rate. This is quite consistent with the fact that the generalization capability of machine learning methods depends on the amount of training examples in general. What makes the problem addressed here interesting is the fact that training data gets scarcer naturally when the distance to the reference centroid increases. Consequently, augmented data techniques could be an interesting approach to increase the generalization capability of the method proposed in this study.

## 5. CONCLUSION

In this paper, LSTM is applied to address the problem of volcano event localization on an end-to-end basis, i.e. without any a priori model relating phase picking with localization estimation. This is particularly interesting because automatic phase picking in volcano signals is a difficult task due to the short distance between the event epicenters and the seismic stations. LSTM was chosen because of its capability to capture the dynamics of time varying signals, and to remove or add information within the memory cell state and model long-term dependencies. The results reported here show that the LSTM based architecture provided a success rate, i.e. an error smaller than 1.0Km, equal to 48.5% with the information provided by only three stations. In contrast, the automatic phase picking delivered very poor success rates equal to 0.3% and 1% with three and up to ten stations respectively. Also, the proposed end-to-end LSTM based method gave a success rate 18% higher than CNN. These results basically corroborate the assumptions that support the choice of the LSTM based strategy and represents a step forward the automatization of some of the tasks in volcano observatories. The combination of the proposed method with data augmentation is proposed for future research.

## 6. ACKNOWLEDGEMENT

The research reported in this paper was funded by grant ANID/FONDEF ID19I-10397. The authors would also like to thank Dr. Ivo Fustos for providing access to Hyposat.

## REFERENCES

Álvarez, I., García, L., Mota, S., Cortés, G., Benítez, C., & De la Torre, Á. (2013). An automatic P-phase picking algorithm based on adaptive multiband processing. IEEE Geoscience and remote sensing letters, 10(6), 1488-1492.

Anantrasirichai, N., Biggs, J., Albino, F., & Bull, D. (2019). A deep learning approach to detecting volcano deformation from satellite imagery using synthetic datasets. Remote Sensing of Environment, 230, 111179.

Audhkhasi, K., Rosenberg, A., Sethy, A., Ramabhadran, B., & Kingsbury, B. (2017). End-to-end ASR-free keyword search from speech. IEEE Journal of Selected Topics in Signal Processing, 11(8), 1351-1359.

Bicego, M., Londoño-Bonilla, J. M., & Orozco-Alzate, M. (2015). Volcano-seismic events classification using document classification strategies. In International Conference on Image Analysis and Processing (pp. 119-129). Springer, Cham.

Bondár, I., Myers, S. C., Engdahl, E. R., & Bergman, E. A. (2004). Epicentre accuracy based on seismic network criteria. Geophysical Journal International, 156(3), 483-496.

Bormann, P., Wendt, S., & Klinge, K. (2013). Data Analysis and Seismogram Interpretation (draft, under review). In New Manual of Seismological Observatory Practice 2 (NMSOP-2) (pp. 1-151). Deutsches GeoForschungsZentrum GFZ.

Canario, J. P., Mello, R., Curilem, M., Huenupan, F., & Rios, R. (2020). In-depth comparison of deep artificial neural network architectures on seismic events classification. Journal of Volcanology and Geothermal Research, 401, 106881.

Cassisi, C., Prestifilippo, M., Cannata, A., Montalto, P., Patane, D., & Privitera, E. (2016). Probabilistic reasoning over seismic time series: Volcano monitoring by hidden Markov models at Mt. Etna. Pure and Applied Geophysics, 173(7), 2365-2386.

Cortés, G., García, L., Álvarez, I., Benítez, C., de la Torre, Á., & Ibáñez, J. (2014). Parallel system architecture (PSA): An efficient approach for automatic recognition of volcano-seismic events. Journal of Volcanology and Geothermal Research, 271, 1-10.

Esposito, A. M., D'Auria, L., Giudicepietro, F., Caputo, T., & Martini, M. (2013). Neural analysis of seismic data: applications to the monitoring of Mt. Vesuvius. Annals of Geophysics, 56(4), 0446.

Esposito, A. M., Giudicepietro, F., Scarpetta, S., & Khilnani, S. (2018). A neural approach for hybrid events discrimination at Stromboli volcano. In Multidisciplinary Approaches to Neural Computing (pp. 11-21). Springer, Cham.

Fatolazadeh, F., Goïta, K., & Azar, R. J. (2020). Determination of earthquake epicentres based upon invariant quantities of GRACE strain gravity tensors. Scientific reports, 10(1), 1-14.

García, L., Alguacil, G., Titos, M., Cocina, O., De la Torre, A., & Benítez, C. (2019). Automatic S-phase picking for volcano-tectonic earthquakes using spectral dissimilarity analysis. IEEE Geoscience and Remote Sensing Letters, 17(5), 874-878.

Gasparini, P., Scarpa, R., & Aki, K. (Eds.). (2012). Volcanic seismology (Vol. 3). Springer Science & Business Media.

Hochreiter, S., & Schmidhuber, J. (1997). Long short-term memory. Neural computation, 9(8), 1735-1780.

Kalkan, E. (2021). An automated P-phase Arrival Time Picker with SNR output (https://www.mathworks.com/matlabcentral/fileexchange/57729-an-automated-p-phase-arrival-time-picker-with-snr-output), MATLAB Central File Exchange.

Kriegerowski, M., Petersen, G. M., Vasyura-Bathke, H., & Ohrnberger, M. (2019). A deep convolutional neural network for localization of clustered earthquakes based on multistation full waveforms. Seismological Research Letters, 90(2A), 510-516.

Krizhevsky, A., Sutskever, I., & Hinton, G. E. (2012). Imagenet classification with deep convolutional neural networks. Advances in neural information processing systems, 25, 1097-1105.

Lara, F., Lara-Cueva, R., Larco, J. C., Carrera, E. V., & León, R. (2021). A deep learning approach for automatic recognition of seismo-volcanic events at the Cotopaxi volcano. Journal of Volcanology and Geothermal Research, 409, 107142.

LeCun, Y., Bengio, Y., & Hinton, G. (2015). Deep learning. nature, 521(7553), 436-444.

Li, B., Chang, S. Y., Sainath, T. N., Pang, R., He, Y., Strohman, T., & Wu, Y. (2020). Towards fast and accurate streaming end-to-end ASR. In ICASSP 2020-2020 IEEE International Conference on Acoustics, Speech and Signal Processing (ICASSP) (pp. 6069-6073). IEEE.

Liu, L., Wang, R., Xie, C., Yang, P., Wang, F., Sudirman, S., & Liu, W. (2019). PestNet: An end-to-end deep learning approach for large-scale multi-class pest detection and classification. IEEE Access, 7, 45301-45312.

Lomnitz, C. (1977). A fast epicenter location program. Bulletin of the Seismological Society of America, 67(2), 425-431.

Malfante, M., Dalla Mura, M., Métaxian, J. P., Mars, J. I., Macedo, O., & Inza, A. (2018). Machine learning for volcano-seismic signals: Challenges and perspectives. IEEE Signal Processing Magazine, 35(2), 20-30.

Masotti, M., Falsaperla, S., Langer, H., Spampinato, S., & Campanini, R. (2006). Application of Support Vector Machine to the classification of volcanic tremor at Etna, Italy. Geophysical research letters, 33(20).

Medsker, L., & Jain, L. C. (Eds.). (1999). Recurrent neural networks: design and applications. CRC press.

Morioka, H., Kumagai, H., & Maeda, T. (2017). Theoretical basis of the amplitude source location method for volcano-seismic signals. Journal of Geophysical Research: Solid Earth, 122(8), 6538-6551.

Mosher, S. G., & Audet, P. (2020). Automatic Detection and Location of Seismic Events From Time-Delay Projection Mapping and Neural Network Classification. Journal of Geophysical Research: Solid Earth, 125(10), e2020JB019426.

Mousavi, S. M., Ellsworth, W. L., Zhu, W., Chuang, L. Y., & Beroza, G. C. (2020). Earthquake transformer—an attentive deep-learning model for simultaneous earthquake detection and phase picking. Nature communications, 11(1), 1-12.

Nicolis, O., Plaza, F., & Salas, R. (2021). Prediction of intensity and location of seismic events using deep learning. Spatial Statistics, 42, 100442.

Pardo, E., Garfias, C., & Malpica, N. (2019). Seismic phase picking using convolutional networks. IEEE Transactions on Geoscience and Remote Sensing, 57(9), 7086-7092.

Pascanu, R., Mikolov, T., & Bengio, Y. (2013). On the difficulty of training recurrent neural networks. In International conference on machine learning (pp. 1310-1318). PMLR.

Perol, T., Gharbi, M., & Denolle, M. (2018). Convolutional neural network for earthquake detection and location. Science Advances, 4(2), e1700578.

Quang, P. B., Gaillard, P., Cano, Y., & Ulzibat, M. (2015). Detection and classification of seismic events with progressive multi-channel correlation and hidden Markov models. Computers & Geosciences, 83, 110-119.

Ross, Z. E., & Ben-Zion, Y. (2014). Automatic picking of direct P, S seismic phases and fault zone head waves. Geophysical Journal International, 199(1), 368-381.

Schmidhuber, J. (2015). Deep learning in neural networks: An overview. Neural networks, 61, 85-117.

Schweitzer, J. (2001). HYPOSAT–An enhanced routine to locate seismic events. Pure and Applied Geophysics, 158(1), 277-289.

Shi, Y., Hwang, M.Y., Lei, X.(2019). "End-to-end Speech Recognition Using a High Rank LSTM-CTC Based Model". ICASSP 2019, Brighton, UK.

Simard, P. Y., Steinkraus, D., & Platt, J. C. (2003). Best practices for convolutional neural networks applied to visual document analysis. In Icdar (Vol. 3, No. 2003).

Staudemeyer, R. C., & Morris, E. R. (2019). Understanding LSTM--a tutorial into Long Short-Term Memory Recurrent Neural Networks. arXiv preprint arXiv:1909.09586.

Stein, S., & Wysession, M. (2009). An introduction to seismology, earthquakes, and earth structure. John Wiley & Sons.

Tan, F., Kao, H., Nissen, E., & Eaton, D. (2019). Seismicity-scanning based on navigated automatic phase-picking. Journal of Geophysical Research: Solid Earth, 124(4), 3802-3818.

Titos, M., Bueno, A., García, L., Benítez, M. C., & Ibañez, J. (2018). Detection and classification of continuous volcano-seismic signals with recurrent neural networks. IEEE Transactions on Geoscience and Remote Sensing, 57(4), 1936-1948.

Titos, M., Bueno, Á., García, L., Zuccarello, L., Álvarez, I., Ibañez, J., & Benítez, C. (2019). Using RNN for automatic detection and classification of volcano seismic signals at Deception Island Volcano. In Geophysical Research Abstracts (Vol. 21).

Trnkoczy, A., Bormann, P., Hanka, W., Holcomb, L. G., & Nigbor, R. L. (2009). Site selection, preparation and installation of seismic stations. In New Manual of Seismological Observatory Practice (NMSOP) (pp. 1-108). Deutsches GeoForschungsZentrum GFZ.

Tsai, K. C., Hu, W., Wu, X., Chen, J., & Han, Z. (2018). First-break automatic picking with deep semisupervised learning neural network. In SEG Technical Program Expanded Abstracts 2018 (pp. 2181-2185). Society of Exploration Geophysicists.

Zeineldeen, M., Zeyer, A., Schlüter, R., & Ney, H. (2020). Layer-normalized LSTM for Hybrid-HMM and End-to-End ASR. In ICASSP 2020-2020 IEEE International Conference on Acoustics, Speech and Signal Processing (ICASSP) (pp. 7679-7683). IEEE.

Zhang, X., Zhang, J., Yuan, C., Liu, S., Chen, Z., & Li, W. (2020). Locating induced earthquakes with a network of seismic stations in Oklahoma via a deep learning method. Scientific reports, 10(1), 1-12.

Zhu, W., & Beroza, G. C. (2019a). PhaseNet: a deep-neural-network-based seismic arrival-time picking method. Geophysical Journal International, 216(1), 261-273.

Zhu, W., Li, X., Liu, C., Xue, F., & Han, Y. (2019b). An STFT-LSTM system for P-wave identification. IEEE Geoscience and Remote Sensing Letters, 17(3), 519-523.

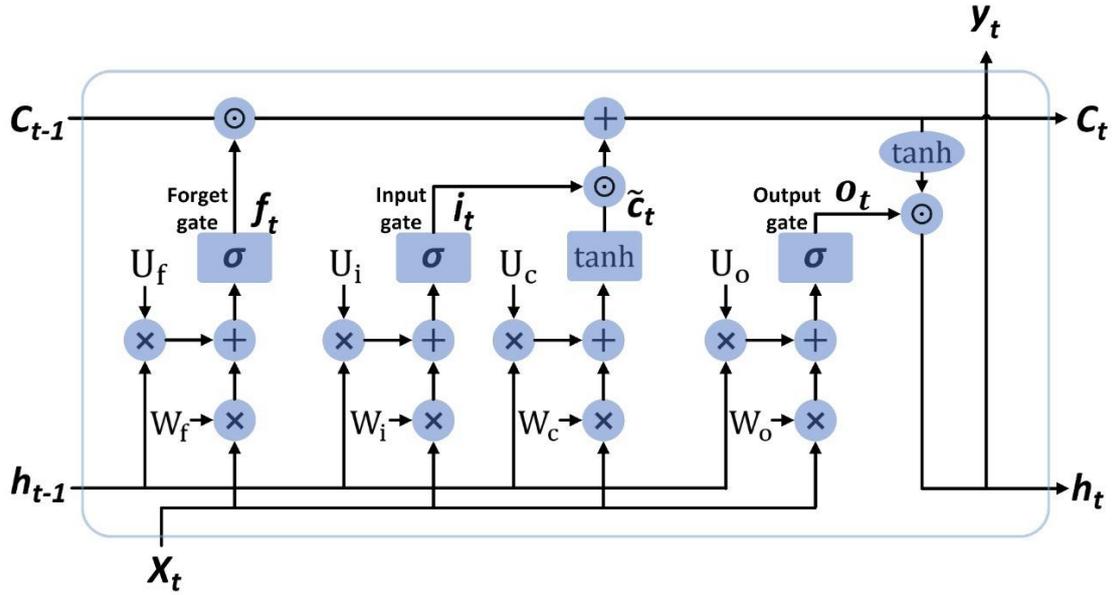

**Figure 1:** LSTM cell and inner block diagram.

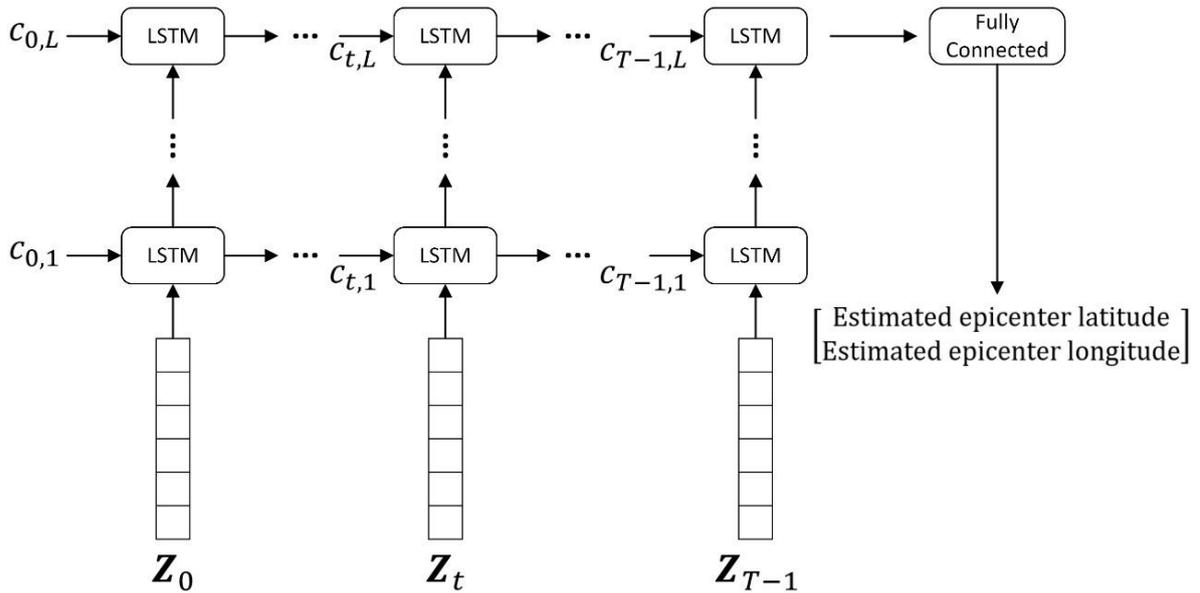

**Figure 2**: The proposed LSTM based architecture for the estimation of volcano event epicenter., i.e. epicenter latitude and longitude. $C_{t,l}$ corresponds to LSTM state at frame t and layer l; $Z_t$ is the input vector corresponding to frame t; T is the number of frames of the input signal; and, L is the number of LSTM layers.

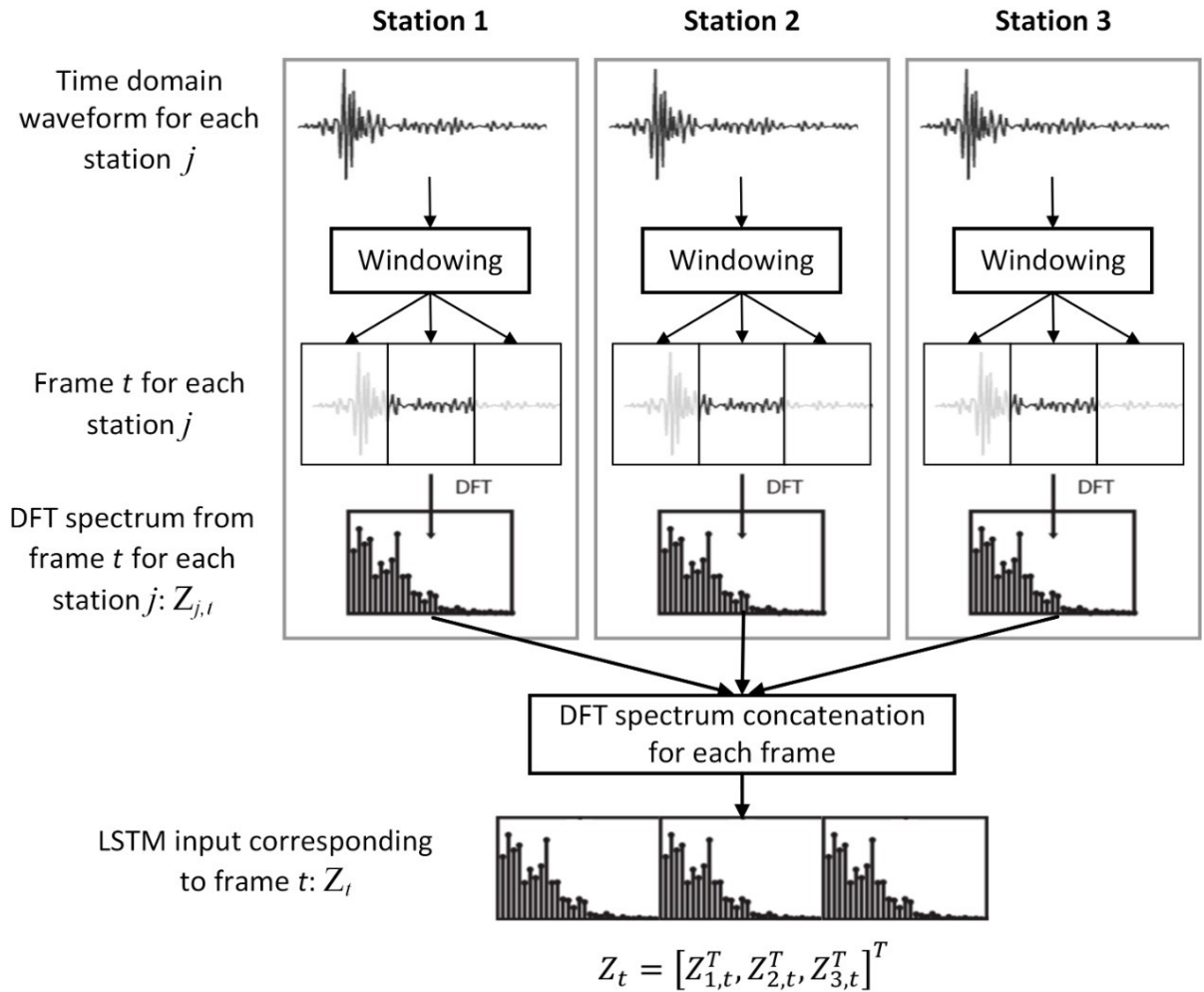

**Figure 3**: Signal pre-processing and feature extraction.

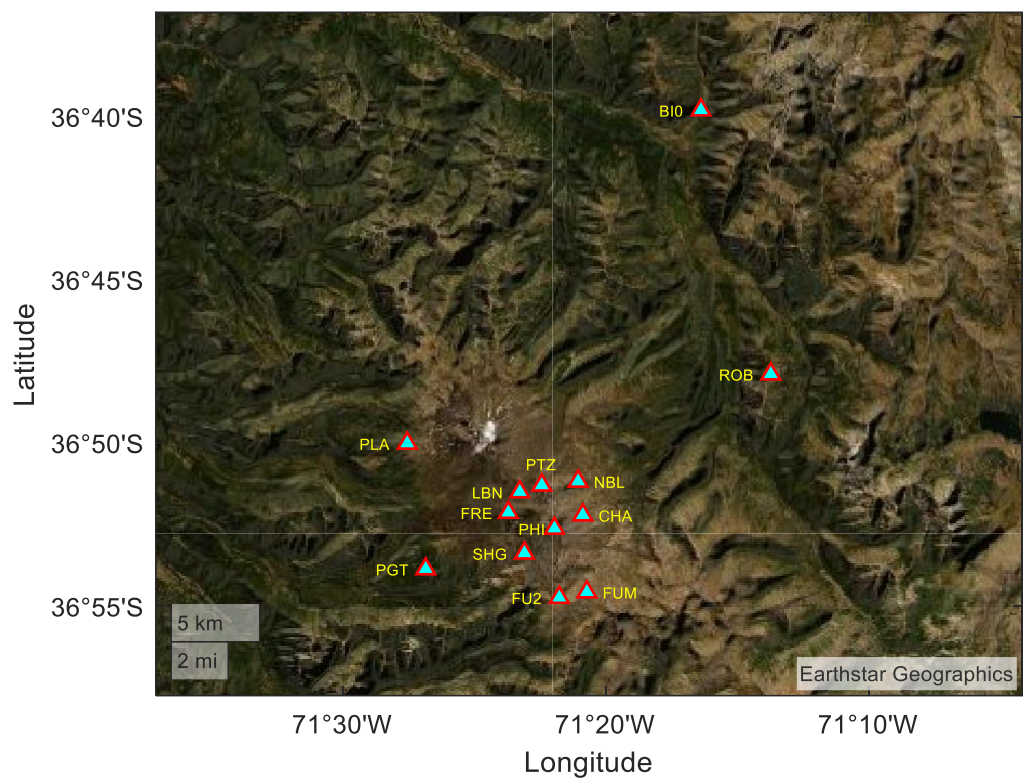

**Figure 4**: Location of Nevados del Chillan volcano. Triangles correspond to the seismic stations.

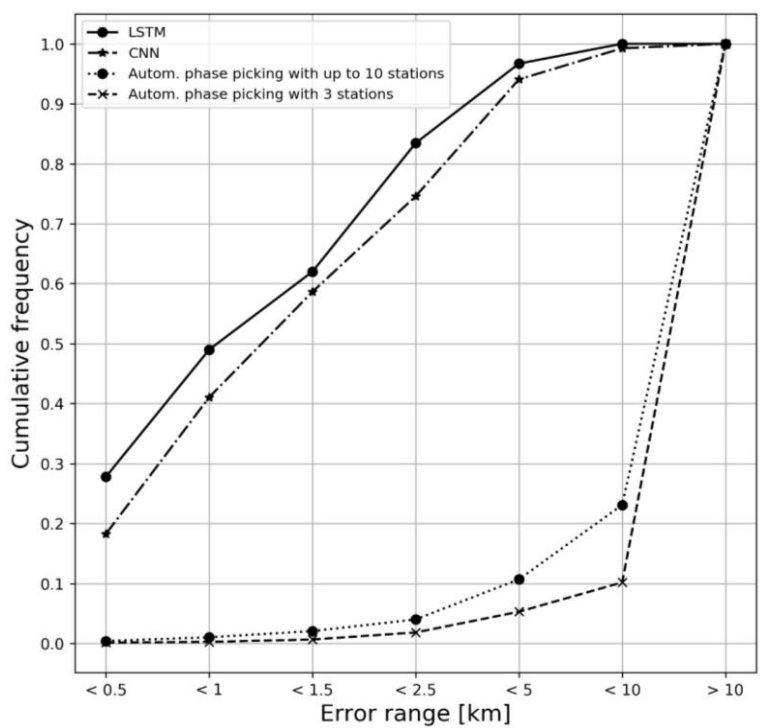

**Figure 5:** Accumulated error with LSTM, CNN and automatic phase picking with three and up to ten stations.

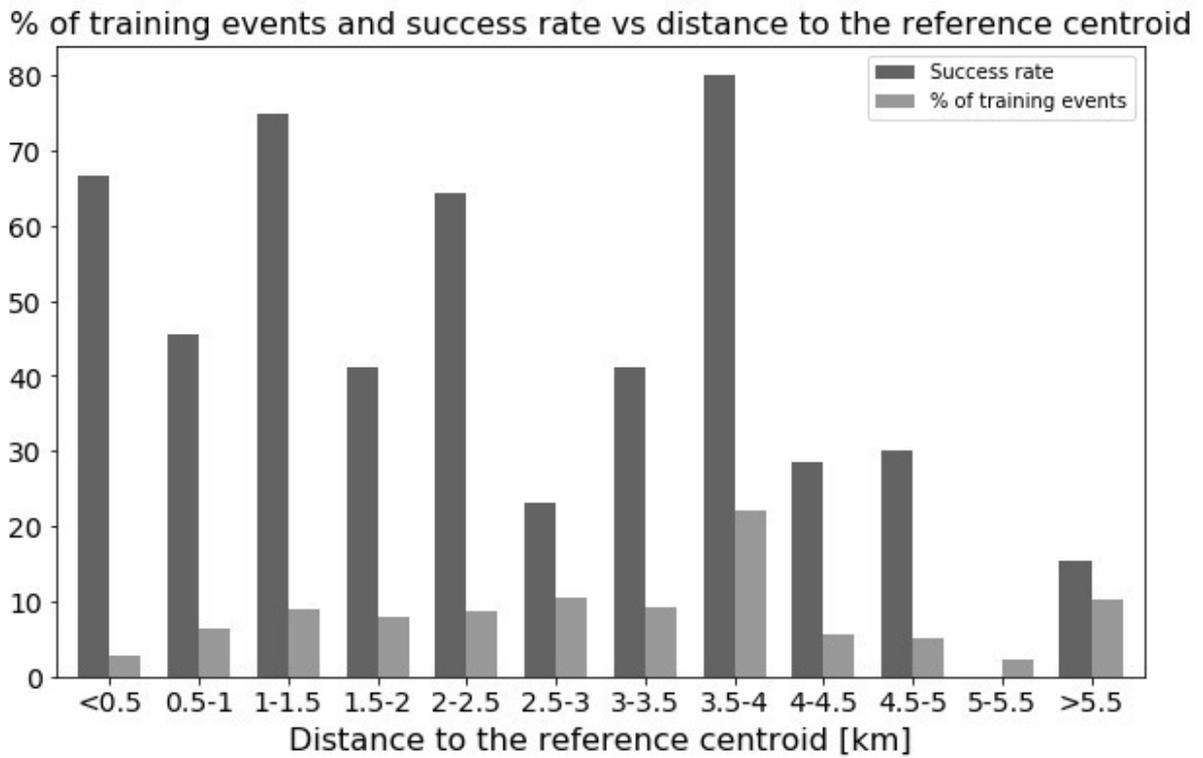

**Figure 6:** % of training events and success rate vs distance to the reference centroid (Km).

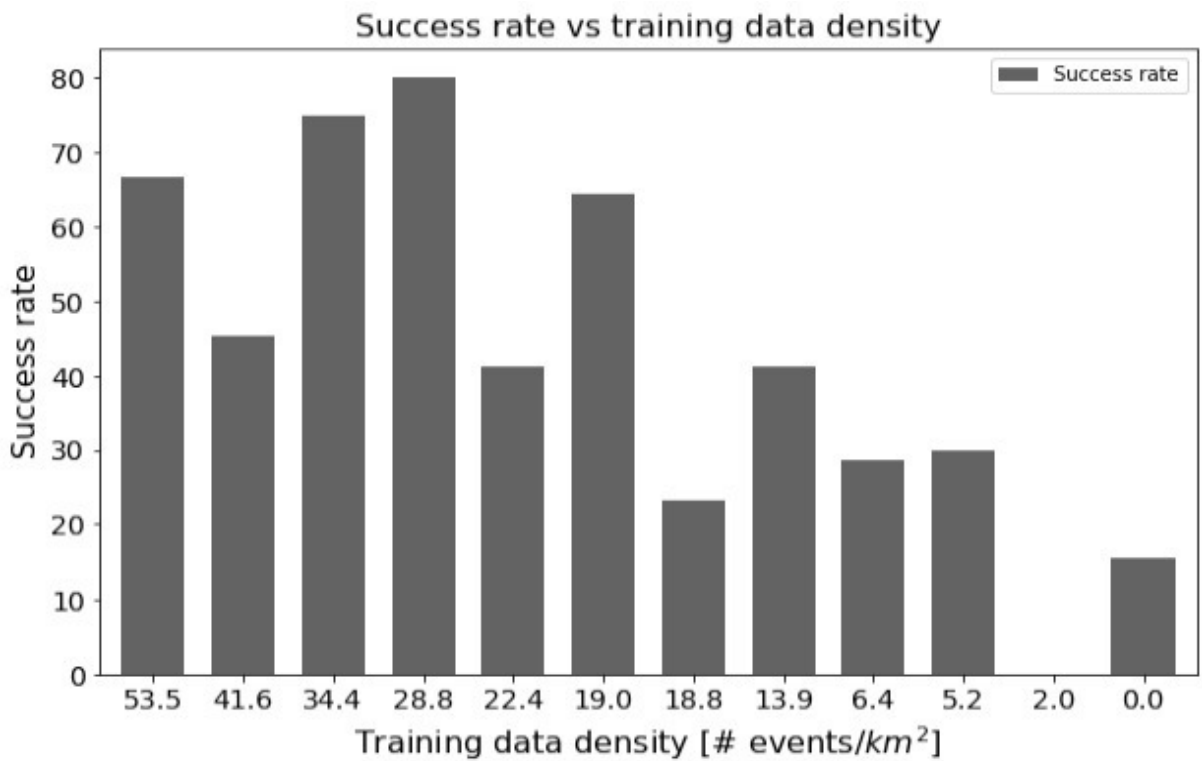

**Figure 7:** Success rate vs. training data density.

16